\documentclass[pra,twocolumn]{revtex4-2}
\usepackage{amsmath}
\usepackage{amssymb}
\usepackage{amsfonts}
\usepackage{relsize}
\usepackage{esint}
\usepackage{bbold}
\usepackage{graphicx}
\usepackage{dcolumn}
\usepackage{bm}
\usepackage{epsfig}
\usepackage{float}
\usepackage[colorlinks=true, citecolor=red, urlcolor=blue, linkcolor=blue]{hyperref}

\usepackage{bookmark}

\usepackage{hyperref}
\usepackage{cleveref}

\begin{document}

\title{Discrepancy between the H-Function and Entropy: Insights into the rigorously
established Boltzmann equation for hard-sphere gases}

\author{Li-Xiang Cen}
\email{lixiangcen@scu.edu.cn}
\affiliation{Center of Theoretical Physics, College of Physics, Sichuan University, Chengdu 610065, China}

\begin{abstract}
By investigating the origin of the evolutionary discrepancy between the H-function
and entropy, we elucidate that the H-function fails to serve as a valid arrow-of-time
criterion in the rigorously derived Boltzmann equation for the hard-sphere gas model
under the Boltzmann-Grad limit.
\end{abstract}

\maketitle

\section{Introduction}

The formulation of the Boltzmann equation stands as a landmark achievement in statistical physics.
It established the foundational paradigm for non-equilibrium statistical physics,
providing a theoretical framework for studying the intrinsic evolution dynamics of
non-equilibrium systems, as well as an effective approach to characterizing their
relaxation toward equilibrium. Constrained by the historical context of its era,
the equation was developed
entirely within the classical statistical mechanics framework, and its embedded
molecular chaos assumption remains a crude approximation for many real
physical systems.
In research related to Hilbert's sixth problem that contributed to
Yu Deng being awarded a Fields Medal, a core achievement is the rigorous derivation of the
Boltzmann equation from Newtonian mechanics--a finding that has reignited scholarly
attention to long-standing historical controversies, including contradictions between
the entropy increase principle
(or H-theorem) and time-reversal symmetry, as well as the Poincar\'{e}
recurrence theorem.

In essence, these ``ancient" controversies stem from the fundamental
tension between the molecular chaos assumption and the reversibility of
mechanical laws. Building on his eponymous equation, Boltzmann introduced
the H-function and derived the H-theorem to explain the directional nature
of system evolution (the so-called arrow of time). This conclusion contradicts
two fundamental mechanical principles for isolated systems: time-reversal
symmetry (giving rise to the Loschmidt paradox) and the Poincar\'{e} recurrence theorem.
Contemporary research in open quantum systems and non-equilibrium quantum
statistics has offered far deeper insights into these ``ancient" disputes:
the molecular chaos assumption neglects statistical correlations between
particle velocity distributions, and its approximate effect is functionally
equivalent to environmental dissipation in real systems. Since no real system
can be perfectly isolated, it is environmental dissipation and decoherence
that give rise to entropy increase and the arrow of time in system evolution.

The paper by Yu Deng, Z. Hani, and Xiao Ma (hereafter DHM) \cite{deng} investigates
the conditions and time domain for the rigorous derivation of the Boltzmann
equation from microscopic dynamics within the mathematical model of hard-sphere
gas under the Boltzmann-Grad (BG) limit \cite{grad}. It extends O. Lanford's 1975
short-time rigorous validity result \cite{lanford} to a long-time limit. This result
has resurrected the aforementioned ``ancient" controversies, prompting considerable
confusion within the academic community.

In this article we will first demonstrate that the rigorously derived Boltzmann equation
for the hard-sphere gas model under the Boltzmann-Grad (BG) limit exhibits
intrinsic anomalous properties. Then we will reveal the marked behavioral
divergence between the H-function and the entropy function during the evolution
of this special mathematical model, that is, the H-function monotonically decreases
while the entropy function remains invariant.
Consequently, the underlying reason for the breakdown of the H-function's
validity as an arrow-of-time criterion in this model will be elucidated.

\section{Anomalous Properties of the Boltzmann Equation for the Hard-Sphere Gas Model
under BG Limit}

The evolution of many-particle interacting systems involves particle distributions
ranging from two-particle up to $N$-particle correlations (with $N$ denoting the total
number of particles), and is in principle governed by the highly complex BBGKY
hierarchy--a chain of $N$ coupled differential-integral equations \cite{BBGKY}. The DHM paper
demonstrates that, under the BG limit, the BBGKY hierarchy for hard-sphere gases degenerates,
in an almost rigorous manner, into the Boltzmann equation within a specified long-time
limit in the work. The paper completes the analysis and evaluation of
the infinite diagrams of the correlation ``molecular" generated by particle
collisions--a mathematically delicate, arduous, and intricate undertaking
that constitutes a remarkable achievement.

From a physical standpoint, the result disclosed by the DHM paper is simple and clear:
the degeneration of the BBGKY hierarchy into the Boltzmann equation implies that
the influence of particle correlations on the system's evolution is entirely negligible.
However, it is worth noting that, unlike the conventional Boltzmann equation
built on the molecular chaos assumption, the rigorous Boltzmann
equation obtained for this mathematical model under the BG limit conceals
an intrinsic anomalous property of the system: its H-function continues to
decrease monotonically, and the system fails to reach thermodynamic
equilibrium within the long-time limit it defines. This property can be
inferred through a modest extension of the DHM derivation: during the system's
evolution, the contributions to the evolution of the distribution function
arising from the recollision of particles that share a history of direct or
indirect collisions are entirely negligible; whereas whenever a ``molecule"
corresponding to a tree-graph structure collides with an independent new particle
or new ``molecule", randomness inevitably drives the H-function to decrease further.
This anomalous property of the system--namely, the failure to attain
thermodynamic equilibrium within its defined long-time limit--was not
underscored in the DHM paper.

Intriguingly, when AI was queried with the question ``whether a hard-sphere gas
model under the Boltzmann-Grad limit, starting from a non-equilibrium state,
can reach thermodynamic equilibrium through inter-particle collisions,"
the response it produced followed precisely this logical chain: a dilute hard-sphere
gas under the BG limit can reach equilibrium through collisions, its microscopic
dynamics converges rigorously to the Boltzmann equation in this limit, and the
H-theorem guarantees convergence toward the Maxwell-Boltzmann distribution.
Subverting this fixed mindset serves as the starting point
for the deduction and resolution of the aforementioned debate of this paper.

\section{Failure to reach equilibrium, or failure to evolve toward it?}

Let us compare the behaviors of the H-function and the entropy in conventional
situations versus that in the current special system. For the Boltzmann equation
established under the molecular chaos assumption:
during evolution, the H-function decreases while the entropy increases;
upon reaching equilibrium, the two attain their minimum and maximum, respectively.
The molecular chaos assumption discards inter-particle correlations, which is
tantamount to forfeiting part of the system's microscopic state
information--the very origin of the temporal asymmetry and irreversibility manifested
in macroscopic evolution. Within this framework, employing either the H-function
or the entropy function as an indicator of the direction of evolution is equivalent.

For the rigorous Boltzmann equation of the hard-sphere gas model under the BG limit,
the H-function decreases monotonically throughout the evolution (till the equation
becomes no longer valid)-but does the entropy
likewise increase in inverse correlation with the H-function? The answer is no.
This can be ascertained from two perspectives: first, the system under consideration
is a rigorously isolated system that obeys the Liouville theorem, and therefore
its entropy remains constant; second, the DHM paper demonstrates that the correlation
terms arising from direct and indirect particle collisions in the system are mathematically
entirely negligible, which implies that the entropy increase induced by discarding
these correlation terms is infinitesimal (this fact reaffirms
that the model cannot attain equilibrium within its defined long-time limit,
for otherwise the entropy increase would have to assume a finite value).

In light of the recognition that the entropy remains constant during evolution,
the hard-sphere gas model under the BG limit, although rigorously governed by the
Boltzmann equation, can only be said to possess a monotonically decreasing H-function;
it cannot be said that the system is evolving toward equilibrium. The Boltzmann
equation holds, yet the H-function--the very criterion for the direction of
evolution-has lost its diagnostic efficacy.

\section{A Subtle yet overlooked distinction between the entropy and the H-Function}

Consider a rigorously isolated gas system initially in a non-equilibrium state.
When two particles with mutually independent probability distributions collide,
the H-function of the system decreases. The underlying physical mechanism is that
the stochastic outcomes of collisions drive the system toward the most probable
distribution--that is, a state in which the velocities obey the Maxwell distribution.
However, the collision process simultaneously establishes statistical correlations
among the particles; when correlated particles collide, the collision
outcomes no longer satisfy the stochasticity assumption, and may in fact cause the
H-function to increase. Consequently, in a rigorously isolated system, the collision
processes among gas molecules drive the H-function to exhibit a dynamically
oscillatory character.

In contrast to the H-function, the definition of the entropy involves the probability
distribution over the ensemble, rather than on the actual distribution of real gas
molecules. If inter-particle collision interactions are entirely neglected,
the ensemble distribution reduces to the particle number distribution over the
energy levels (i.e., the momentum-space distribution) of free particles.
The entropy is then completely equivalent in nature to the H-function.
Once collision interactions are, however, taken into account, every element of the
ensemble--that is, every microstate--is an eigenstate of the system's Hamiltonian,
which has already incorporated the inter-particle collision interactions. Collisions
therefore induce no transitions among the distinct states of the ensemble, leave the
probability distribution over the ensemble unchanged, and consequently leave the
entropy of the system unaltered. This is precisely the conclusion of the
Liouville theorem.

The subtle distinction between the H-function and the entropy function outlined
above has hitherto not attracted widespread attention. Intriguingly, this distinction
is magnified during the evolution governed by the rigorous Boltzmann equation
of the hard-sphere gas model under
the BG limit: the entropy remains constant, while the H-function decreases
monotonically over its defined long-time limit.

\section{The H-Function versus the entropy: Which sets the direction of evolution?}

The answer is, of course, the entropy function. Entropy increase represents
the loss of information, the growth of disorder, and the irreversibility and
temporal directionality of physical processes. This point has remained unshaken
throughout the foregoing discussion. In contrast to the standpoint adopted in
those ``ancient" debates, modern physics regards entropy increase as being induced
by the dissipative effect of the environment--be it the vacuum background field.
Yet how is one to understand the failure of the
H-function here and the striking contrast it exhibits? One may conceive the
following scenario: the collision process causes the ``molecular" clusters
corresponding to tree-graph structures to grow ever larger; when the time
exceeds the long-time limit defined in the DHM paper, these clusters will
eventually engulf all the remaining independent particles and ``molecules."
The Boltzmann equation is then broken and particle collisions in
the subsequent evolution will cause the
H-function to change in the opposite direction--to increase monotonically.
In other words, throughout the evolution of this isolated system, the
H-function will exhibit dynamical variation still, but with extremely long
oscillation periods and extremely large oscillation magnitudes.
This furnishes an interpretation of the Poincar\'{e} recurrence theorem
for isolated systems.

\section{Final remark}

For a dilute gas system initially obeying the molecular chaos assumption,
the Boltzmann equation emerges naturally as an exact description for the system's dynamics
throughout a very short initial time interval. Mathematicians have performed
meticulous research about the temporal extent of this exact validity regime.
Physically, this time window corresponds to the buildup process
of particle correlations: initially independent particles develop correlations
through collisions, but collisions among already correlated particles have not
yet occurred, or their frequency remains negligible in its impact on the global
distribution of the system.
Even if this timescale can be greatly extended in specific
mathematical models (with DHM demonstrating a time interval scaling as the double
logarithm of the total particle number $N$), such extension does not imply irreversible
evolution of the system. In fact, the evolution over this initial time interval does not
reflect the original physical connotations of the Boltzmann equation.
Rather, it is the discarding of correlations in the subsequent evolution
(an effect analogous to decoherence induced by environmental dissipation,
as noted at the outset of this paper) that endows the
system's evolution with entropy increase and the arrow of time.

\section*{ACKNOWLEDGMENTS}

This work was supported by the National Natural Science Foundation of China (Grant No. 12471443).

\end{document}